\documentclass[11pt]{article}

\usepackage[letterpaper,margin=1in,headheight=16pt]{geometry}
\usepackage[T1]{fontenc}
\usepackage{lmodern}
\usepackage{graphicx}
\usepackage{booktabs}
\usepackage{enumitem}
\usepackage{titlesec}
\usepackage{fancyhdr}
\usepackage{amsmath}	
\usepackage{amssymb}	
\usepackage[square,numbers]{natbib}
\usepackage[most]{tcolorbox}

\newcommand{\apjl}{Astrophysical Journal Letters}
\newcommand{\apjs}{Astrophysical Journal Supplements}

\newcommand{\aap}{Astronomy \& Astrophysics}

\usepackage{xcolor}
\definecolor{genesisgold}{HTML}{B5771A}
\definecolor{genesisink}{HTML}{10131F}
\definecolor{genesisgray}{HTML}{6B7386}

\usepackage[colorlinks=true,linkcolor=genesisgold,citecolor=genesisgold,urlcolor=genesisgold]{hyperref}

\titleformat{\section}{\normalfont\large\bfseries\color{genesisink}}{}{0pt}{}
\titlespacing*{\section}{0pt}{1.3em}{0.45em}

\fancypagestyle{genesisfirst}{%
  \fancyhf{}%
  \fancyhead[L]{\raisebox{-3pt}{\includegraphics[height=13pt]{genesis-logo}}\hspace{4pt}\small\textsc{\textcolor{genesisgold}{JWST Genesis Initiative}}}%
  \fancyhead[R]{\small\textcolor{genesisgray}{MCT Concept White Paper}}%
  \fancyfoot[C]{\small\thepage}%
}

\usepackage{tikz}
\newcommand{\gq}{\leavevmode\tikz[baseline=-0.6ex]{\fill[genesisgold] circle(0.65ex);}\hspace{0.2em}}

\begin{document}
\thispagestyle{genesisfirst}

 \vspace{-1mm}
{\fontsize{16}{19}\selectfont\bfseries\color{genesisink}
Why a Multi-cycle Treasury Program is essential to maximize JWST’s scientific contributions to galaxy formation\par}
\vspace{-2mm}

\begin{tcolorbox}[
  enhanced, breakable,
  colback=genesisgold!8,
  colframe=genesisgold,
  boxrule=0.8pt,
  arc=4pt,
  boxsep=0pt,
  top=2pt, bottom=4pt,
  left=0.6em, right=0.6em
]
\noindent\raisebox{-0.3\height}{\includegraphics[height=1.5em]{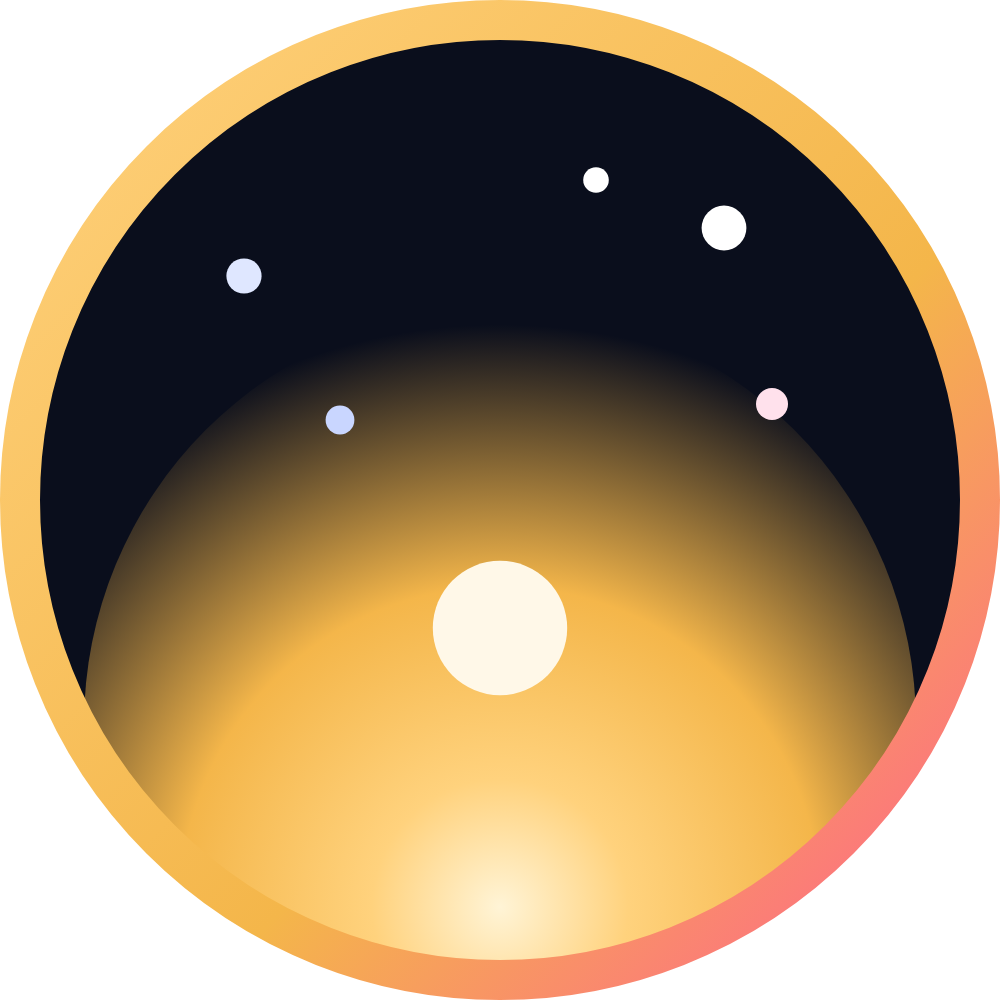}}\hspace{0.5em}%
\emph{\textbf{The Genesis Initiative} is an extragalactic-community-wide effort to coordinate collaboration in response to the STScI MCT concept white paper (WP) call. Information and WP titles are on the \href{https://www.jwst-genesis.com}{Genesis Initiative website}.  Authors and co-signers of this WP are on page 6. Two companion Genesis Initiative theory white papers are available \citep{genesistheory1,genesistheory2}.}
\end{tcolorbox}

\vspace{1.0em}

 \vspace{-10mm}
\section{Introduction}
\vspace{-2mm}

 Two of the James Webb Space Telescope's (JWST) primary science goals are ``discovering and characterizing the first stars and galaxies'', and ``understanding how these objects evolved into the galaxies we see in the nearby Universe'' \citep{gardner2006}. These questions -- when the first galaxies emerged from the cosmic dark ages, how the first massive black holes formed, and how the heavy elements necessary for life were forged in the first stars —- together mark the \textbf{\emph{\textcolor{genesisgold}{genesis}}} of our own Milky Way, and of the conditions that made life possible.

The fundamental importance of these questions has been called out by the past three Astronomy and Astrophysics decadal surveys, and formed the primary motivation for investing decades of effort and billions of dollars to making the JWST mission a reality. Most recently, the Astro2020 decadal identified ``Cosmic Ecosystems: Unveiling the Hidden Drivers of Galaxy Growth'' as a priority science area, stating: ``An essential milestone on the pathway to revealing these drivers is a complete understanding of the formation and buildup of galaxies and their structures, stellar populations, metals, and central black holes over cosmic time. JWST is poised to revolutionize this subject by imaging and identifying galaxies out to some of the first generations during the epoch of reionization, and by obtaining deep redshifted ultraviolet–visible spectra for thousands of galaxies spanning the period from reionization to the present.'' \citep{NASEM2023_Pathways}. Other key questions identified by the Astro2020 science panels that JWST can address include: 1) What are the seeds of supermassive black holes and how do they grow? 2) How did the intergalactic medium and the first sources of radiation evolve from cosmic dawn through the epoch of reionization?  3) How do gas, metals, and dust flow into, through, and out of galaxies?   4) How do supermassive black holes form, and how is their growth coupled to the evolution of their host galaxies? \citep{NASEM2023_Pathways_Ch2} 

JWST’s performance has been even better than expected, and the Universe has also cooperated by disclosing a bounty of unexpected discoveries. Despite the associated excitement, these discoveries have posed as many new questions as they have answered, and major goals of the JWST mission arguably remain un-achieved. A \emph{uniform, statistically controlled legacy dataset is now required} to transform these discoveries into a quantitative physical understanding of early galaxy formation. Here we argue that this requires a coordinated and ambitious \underline{community-led} strategy, demanding a significantly larger time allocation than has been approved through the standard TAC process to date. Over the past five years, the JWST user community has gained substantial expertise in designing observing strategies and in interpreting the resulting data. \textbf{\emph{Now is the time}} for JWST to embark on the next phase of scientific discovery, by enabling these kinds of larger efforts.
\vspace{-4mm}

\section{Lessons from the Hubble Space Telescope}
\vspace{-3mm}

Large, ambitious extragalactic surveys with space telescopes have a long and successful legacy, beginning in 1995 when HST devoted ten consecutive days of telescope time to an apparently blank patch of sky, creating the iconic Hubble Deep Field \citep{williams96}. While this program was considered a significant risk, and arguably never would have survived the standard peer-review TAC process, these observations revealed a universe teeming with galaxies at every stage of evolution, fundamentally changing our understanding of galaxy formation.  The HDF pioneered an entirely new mode of “blank-field” astronomy, which is now the dominant observational paradigm for early universe discovery.  
Major advances in our empirical understanding of galaxy formation over the past three decades trace back to this lineage: from the HDF \citep{williams96} (done in {\it HST} Cycle 5!) to the Hubble Ultra Deep Field \citep{beckwith06} and Great Observatories Origins Deep Survey (GOODS, a GO program; \citep{giavalisco04}), to the Hubble Frontier Fields \citep{lotz17}, and many in between.
These programs were made possible through Director's Discretionary Time (DDT). Their success led to the first HST Multi-Cycle Treasury (MCT) program call to enable programs far beyond what the standard TAC process could accommodate. This call resulted in three transformative programs: CANDELS \citep{grogin11,koekemoer11}, the largest HST survey ever at the time, which mapped galaxy evolution across five legacy fields over 902 orbits; CLASH (524 orbits), which used cluster lensing to probe the high-redshift universe \citep[][paving the way for the HFFs]{postman12}; and PHAT (828 orbits), which resolved stellar populations across a third of the Andromeda galaxy \citep{dalcanton12}. DDT contributed roughly a third of the $\sim$2250 allocated orbits, with the remainder drawn from the GO pool. A second MCT call in 2023 yielded two programs now  underway \citep{clutch,stela}.\looseness=-2

Figure 1 shows the distribution of HST GO program sizes, which are predominantly $\leq$200 orbits.  Only Treasury/MCT and DDT programs exceed this threshold, accounting for all programs which exceed 400 orbits.  Among these are the most transformative studies in galaxy evolution history which have continuously pushed back our redshift horizon to $z \sim$ 4 (HDF), $z \sim$ 6 (e.g., GOODS and COSMOS \citep{scoville07}), and $z \sim$ 10 (e.g., CANDELS and CLASH). These programs are highly impactful even 1--2 decades after survey completion, providing legacy data in the key fields which continue to be observed with JWST and other facilities today. GOODS and CANDELS are generating $\sim$10 and 5 citations/orbit/year even today.  The gold histogram indicates the JWST GO program distribution -- similar to HST, the TAC-gated GO programs rarely exceed 300 hr.
\vspace{-4mm}

\begin{figure}[t!]
\begin{minipage}{0.56\textwidth}
\hspace{-3mm}
\includegraphics[width=\textwidth]{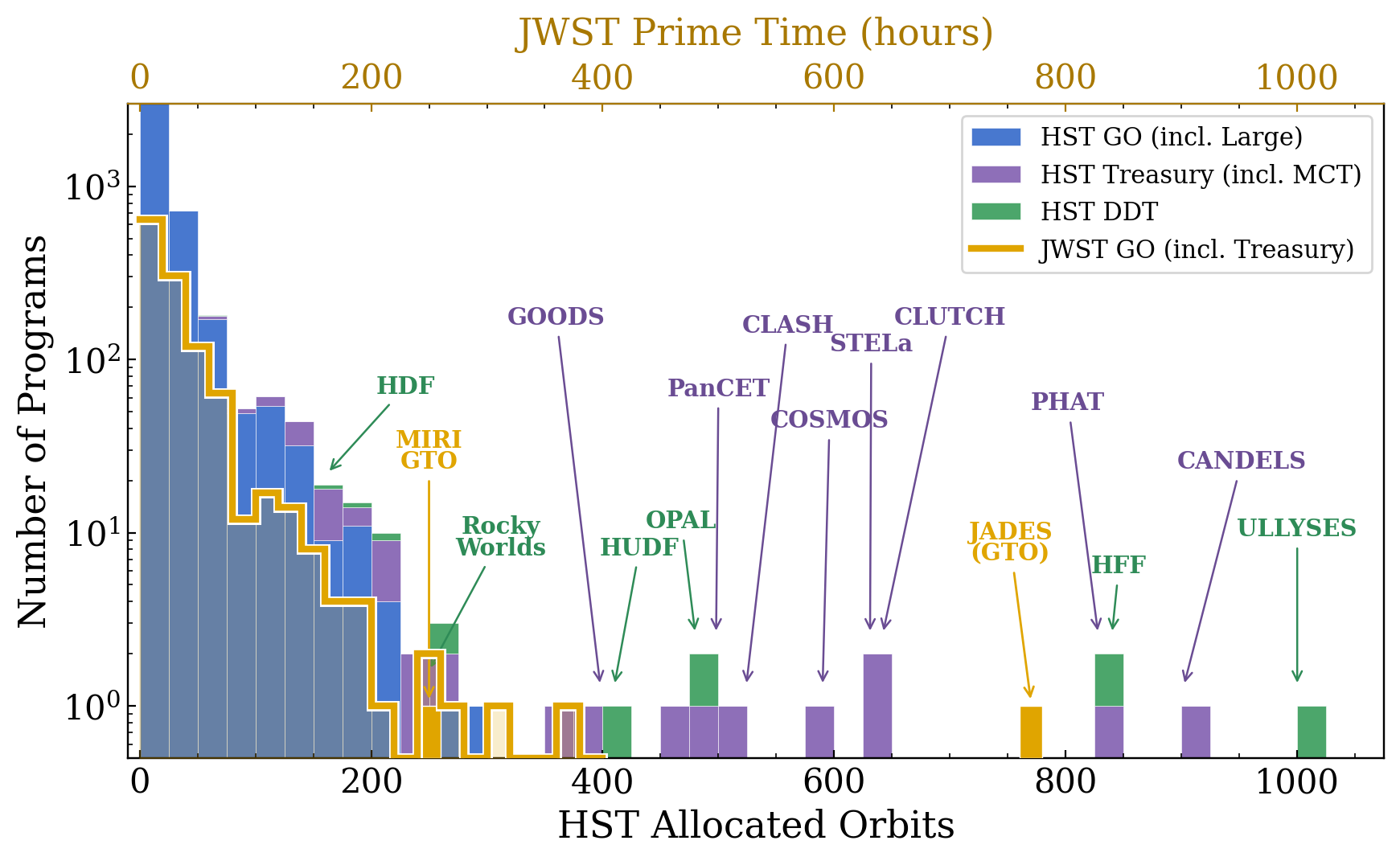}
\end{minipage}
\begin{minipage}{0.43\textwidth}
\caption{Distribution of approved GO program sizes for HST (in orbits; from the TALL database) and JWST (in hours; from STScI approved-program GO listings).  For HST, we indicate Treasury programs including MCT in purple, and DDT programs in green \citep{Reid2026}.  We also highlight the JWST GTO programs JADES (NIRCam$+$NIRSpec) and MIDIS (MIRI) programs.\looseness=-2}
\label{fig:growth}
\end{minipage}
\vspace{-6mm}
\end{figure}

\section{The Galaxy Formation Landscape and Key Unanswered Questions}
\vspace{-3mm}

The observable properties of galaxies are forged by a complex suite of multi-scale and multi-physics processes. Scaling relations (correlations between observable properties such as mass, internal velocity, size, star formation rate [SFR], and metallicity) encode these processes and are essential for constraining theories of galaxy formation. Over the past several decades, HST, Spitzer, and ground-based observations pushed our view of galaxy evolution ever farther back in cosmic time, revealing the evolution of these fundamental scaling relations and the emergence of the Hubble sequence.\looseness=-2

Just prior to JWST's launch, the redshift frontier had been pushed back to $z \sim10$ \citep[e.g.,][]{oesch16,Finkelstein2022}.  
Empirical extrapolations from these observations as well as the general consensus of theoretical models predicted that the cosmic SFR density and the number density of UV-bright galaxies would continue to decline towards even higher redshifts \citep[with the exact rate of decline debated; e.g.,][]{bouwens16,Oesch:2018,mcleod16}, forecasting that JWST should see few if any UV luminous galaxies at $z>10$ in the initial rather small surveys. Instead, JWST has transformed our understanding of the very early universe by revealing that galaxies grow surprisingly fast and furiously at cosmic dawn, reaching unexpectedly high luminosities and masses, at earlier times than most pre-launch theoretical and empirical models predicted \citep[e.g.,][]{Finkelstein2023,Harikane2023,Leung2023,Adams2024,Donnan2024}. Moreover, the exquisite capability of the NIRSpec instrument onboard JWST has not only confirmed the redshifts of many of these high-$z$ objects \citep[e.g.,][]{Curtis-Lake2023,ArrabalHaro2023}, but has begun to reveal their internal physical properties. These studies suggest that early galaxies have experienced significant chemical enrichment \citep{Bunker2023,calabro24}. Some have elemental abundance ratios and ISM properties that are significantly different from those of their lower redshift counterparts, and perhaps reflective of early star-cluster formation \citep[e.g.,][]{Cameron2023,Topping2024}. 

The highest-redshift spectroscopically confirmed galaxies identified with JWST to date are at redshifts of $z\simeq 14$ \citep{Carniani2024,Naidu2025}, and there are photometric candidates with redshifts as high as $z \gtrsim$20 \citep[e.g.][]{Perez-Gonzalez2025,Gandolfi:2025,Castellano2025}. 
The inferred star formation histories and metallicities of the $z\simeq 14$ objects strongly imply that they should have detectable progenitors at $z\gtrsim 15$ \citep{Ferrara2024}. Although these objects are apparently not composed of Pop III stars, JWST has found tantalizing, though hotly debated, hints of possible pockets of Pop III star formation at lower redshifts \citep[e.g.,][]{Morishita2025,maiolino2026,Ubler26}. Lyman-$\alpha$ emission has been discovered out to $z\sim 13$ \citep{Witstok26}, indicating that regions of the Universe were already re-ionized at these surprisingly early epochs, and painting a picture of highly patchy reionization. 

Another major discovery area is the study of early black hole growth at $z>6$, via the measurement of Balmer emission lines tracing the AGN broad-line region \citep[e.g.,][]{harikane-agns,Taylor2024}. 
In addition, JWST has discovered a mysterious population of objects termed ``Little Red Dots'' \citep[LRDs;][]{Matthee2024}, which have ultra-compact morphologies and unusual ``V-shaped'' spectral energy distributions, and are more common at $z>4$ \citep{Kocevski2024,kokorev24}.  The majority of LRDs exhibit very broad emission lines \citep{greene24}, leading to the interpretation that they are powered by supermassive black holes which may be overmassive relative to their host galaxies \citep[e.g.,][]{pacucci2023,gupta26}, but they are generally undetected in the X-ray and radio \citep[][]{furtak2024,setton2025}, unlike traditional AGN. Their red nature is (typically) not linked to the presence of large amounts of dust \citep{perezgonzalez2024,Casey2025}, complicating the interpretation in terms of the canonical AGN engine structure. Their physical nature remains highly debated, with interpretations ranging from accreting black holes with unusual accretion and obscuration properties \citep{Inayoshi2025} to alternative stellar-powered scenarios, including supermassive stars \citep{nandal2026,perezgonzalez2026,kokorev2026}, star clusters \citep{chisholm2026}, and quasi-stars \citep{naidu2026,begelman26}.\looseness=-2

The ability to perform continuum and absorption-line spectroscopy of massive galaxies at $z\geq 4$ has revealed the existence of a surprisingly large population that apparently assembled and quenched their star formation very early and rapidly \citep{carnall24,glazebrook24}. Some of the tension with current simulations may be relieved by appealing to stellar population models with more realistic chemical abundance patterns \citep{degraaf25}, spatially resolved spectroscopy \citep{cutler26}, and more sophisticated modeling of early star formation and AGN feedback \citep{Somerville2025,Farcy2025}, but just how much remains to be seen. Furthermore, JWST imaging has revealed the emergence of well-developed structural and morphological features like bars, spheroids, and disks, earlier in cosmic history than previously thought \citep[e.g.][]{Huertas-Company2025}.

These discoveries lead to several open questions that JWST can uniquely address:
\gq When and how do the \underline{\textit{earliest galaxies}} first emerge from the cosmic
dark ages, and which processes govern their star formation?
\gq Can we identify and characterize primordial \underline{\textit{Population III}} stars during their lives or
their explosive deaths?
\gq What do the unusual elemental \underline{\textit{abundance patterns}} reveal about the
formation modes of the first galaxies?
\gq When and how do the first \underline{\textit{massive black holes}} form, what are their
masses, and what is their connection to the mysterious population of LRDs?
\gq How do \underline{\textit{massive galaxies}} form and quench so early and rapidly?
\gq When do the \underline{\textit{scaling relations}} that encode the physics regulating
galaxy evolution first emerge, and how do they evolve across the first few billion years?
\gq When and how did \underline{\textit{reionization}} occur, and what were the powering sources?
\vspace{-4mm}

\section{The Limits of Existing Programs to Answer These Questions}
\vspace{-3mm}

Over the course of the first five cycles of JWST observations, inclusive of Early Release Science (ERS \citep{finkelstein25,Castellano2022}), Guaranteed Time Observation (GTO \citep{eisenstein26,midis,canucs,ganifs}), and GO programs, the galaxy formation community has employed many of the telescope’s key observational modes. These include large multi-band NIRCam imaging surveys, utilizing both broad and medium-band filters, undertaken in well-studied legacy fields (e.g., JADES, CEERS, PRIMER, COSMOS-Web, MINERVA) and in lensing cluster fields (e.g., GLASS-ERS, CANUCS, UNCOVER, GLIMPSE, VENUS), while multi-epoch imaging campaigns have enabled high-redshift time-domain astrophysics (e.g., NEXUS, VENUS, PEARLS-NEP). Spectroscopic survey programs have utilized the NIRSpec MSA (e.g., JADES, CEERS, CAPERS, JWS3, RUBIES) and NIRCam grism (e.g., FRESCO, COSMOS-3D, CONGRESS) to assemble large samples. Smaller samples have been observed with NIRSpec and NIRISS to greater depths (e.g., SPURS, DIVER, DAJE, NGDEEP), and with the NIRSpec IFU to spatially map high-redshift galaxies (e.g., GA-NIFS, ALPINE-CRISTAL, REBELS-IFU, BlackTHUNDER). MIRI imaging has been obtained in parallel for several NIRCam programs (e.g., CEERS, PRIMER, COSMOS-Web, MINERVA) and prime in the HUDF (e.g., MIDIS, SMILES, MIRAGE, MEOW), whereas MIRI spectroscopy has yet to be widely deployed over large samples. 

Large quantities of data from ERS, Large, and GTO programs suitable for high-redshift analyses are now publicly available, 
and have been reduced into a science-ready state in archives such as the DAWN JWST Archive \citep{heintz25}. Public imaging data from multiple surveys have been combined to produce powerful statistical samples tracing the evolving galaxy UV luminosity function at $M_{UV}$ $\leq -$18 to $z \sim$ 15 (and even, tentatively, to $z \sim 18$; \citealp[][and references therein]{fontana26}), and the galaxy stellar mass function at $M_* \geq 10^8\,M_{\odot}$ to $z \sim 9$ (and even, tentatively, to $z \sim 13$; e.g., \citealp{weibel2024,harvey2025}).
Lensing enables the extension of the UV luminosity function $\sim$2 magnitudes fainter at $z \sim$ 15 \citep{chemerynska2026,kokorevglimpse}. Spectroscopic emission-line measurements (e.g., H$\alpha$, H$\beta$, [OIII], [OII], [NeIII]) out to $z \sim 10$ allow us to constrain the evolving fundamental baryonic scaling relations among galaxy SFR, stellar mass, metallicity, and chemical abundance patterns; measure the ionizing efficiency and burstiness of galaxy stellar populations; and reveal strikingly different relations between the masses of supermassive black holes and their host galaxies, compared to observations in the local universe.\looseness=-1

Despite significant progress in uncovering the first billion years of galaxies with JWST enabled by this wealth of data, we identify several fundamental limitations of the currently available data:
\vspace{-3mm}

\begin{itemize}[leftmargin=*]


\item The surveyed imaging area is too small at the depth required to uncover a robust $z>$15 photometric sample for spectroscopic follow-up (the deepest imaging [NGDEEP+MIDIS \citep{Leung2023,Perez-Gonzalez2025}] reaches m$_{lim} \sim$ 31 over $\mathcal{O}$(1 arcmin$^2$)). Furthermore, the sample of galaxies at $z>$10 with robust stellar masses based on MIRI rest-optical data is insufficient for determining the dominant physics behind the observed UV luminosity function evolution (bursty star-formation histories, enhanced star-formation efficiency; \citep{Donnan:2025}).
\vspace{-1mm}

\item Despite efforts at photometric pre-selection \citep{fujimoto2025,trussler2026} and spectroscopic observations of low-mass systems at $z>6$ \citep{Morishita2025,maiolino2026}, we still lack any confirmed Pop III systems, as the current upper limits on metal-line fluxes do not reach the Pop III regime.
\vspace{-1mm}

\item There is no consensus regarding the underlying physical model for LRDs. Ongoing debate pertains to the mass scale of their SMBHs (e.g., \citep{greene2026}), or if LRDs even contain BHs (e.g., \citep{chisholm2026,perezgonzalez2026}). Key problems include the heterogeneity of LRD selection criteria, and the small, biased and incomplete samples with high-quality spectra. 

\vspace{-1mm}


\item Simulations \citep[e.g.][]{volonteri2025,koudmani2026} predict that disentangling BH seeding and accretion models will require pushing to either higher redshifts ($z\gtrsim 9$) or lower-mass galaxies than probed by current samples \citep{geris2026}. Uncertainties in single-epoch SMBH estimates must also be surmounted \citep{pacucci2023,shen2024,lupi2024,rusakov2026}.
\vspace{-1mm}

\item Although large photometric samples of apparently quiescent galaxies have been assembled within the first 2 billion years of cosmic time ($z>3$), the number with deep spectroscopic measurements \citep{glazebrook24,carnall23,degraaf25} remains too incomplete to systematically constrain quenching models. 
\vspace{-1mm}

\item Spectroscopic samples at $z\geq4$  have several critical limitations: (1) they are incomplete and biased towards higher than average sSFR values at low masses ($M_*<10^{8.5} M_{\odot}$); (2) the subset of galaxies with sufficient depth for detecting multiple emission lines is still too small to perform robust evolutionary studies of galaxy scaling relations across multiple redshift bins at $z>4$ \citep[e.g.][]{curti24,isobe26}; (3) there is no comprehensive $R\gtrsim1000$ spectroscopic sample with sufficient statistics, which was also selected uniformly from NIRCam imaging \citep[cf. JADES][]{bunker24,curtislake26}. 
Such uniform selection is essential for comparison with galaxy formation simulations.
\vspace{-1mm}




\item Ultradeep spectroscopy has proven particularly powerful for elucidating the origin of multiple chemical elements (C, N, O, Fe) in the earliest galaxies, the nature of massive stars, and the central regions of LRDs \citep[e.g.,][]{Topping2024,Berg2026,Marques-Chaves2026,Tang2026}, yet the sample of distant galaxies with such coverage ($N<100$) is too small for robust population studies. 
\end{itemize}
\vspace{-8mm}

\section{The Time is Now for a JWST MCT-Scale Investment}
\vspace{-3mm}

While JWST is fully capable of acquiring the ambitious datasets needed to answer these questions, the distribution of approved JWST GO programs across Cycles 1-5 tells a clear story: the largest extragalactic GO programs (NEXUS [368 hr], VENUS [303 hr], and COSMOS-Web [255 hr]) represent the practical ceiling of what TACs are willing to approve, and several of these required multi-cycle strategies (both for scheduling and science reasons). The science described above demands larger time allocations: $\sim$ 1000-2000 hr. As discussed in many companion white papers, reaching $z >$ 15 requires NIRCam imaging depths $\gtrsim$ 1 mag deeper than existing data ($\sim$100--150 hr/filter in a deep blank or lensed field to reach $m_{\rm lim} \sim$ 32, or $m_{\rm lim} \sim$ 30 over a substantially wide field). A spectroscopic survey capable of robustly constraining galaxy scaling relations dictates thousands of uniformly selected targets, across a wide dynamic range in source luminosity, requiring long integration times.  The earliest black holes, (relatively) evolved stellar populations, and dust-obscured systems require MIRI imaging depths and areas significantly exceeding what has been achieved (tens of hours per filter over many pointings).   Time domain science requires cadenced observations that are structurally impossible within a single-cycle program.   One or more unified MCT programs will optimize survey design to most effectively use critical observing resources (such as the MSA) with uniform and well characterized selection functions, which are crucial for statistical studies and for addressing multiple science goals, as demonstrated by previous legacy surveys such as the SDSS.\looseness=-1

While existing programs provide invaluable constraints, their heterogeneous selection criteria, depths, and observing strategies limit the construction of statistically complete and uniform samples. A comprehensive JWST MCT program would provide the scale, homogeneity, and multi-cycle continuity needed to address these interconnected science goals \citep[e.g.][]{finkelstein_mct,jff_survey,fontana_mct,harikane_mct}. Such a program would naturally yield high-level data products, catalogs, and tools for comparison with theory.

A suite of extragalactic MCT programs is time-critical: as JWST completes its five-year prime mission, most instruments are still at peak performance, but there are modes that are at risk or already in reduced efficiency (e.g., longer wavelength MIRI, NIRSpec MSA growing stuck shutters). MCT programs over the next few cycles will maximize synergy with contemporary and upcoming facilities such as ALMA, Euclid, Roman, Rubin, SKA, PRIMA, and LISA; identify high-value science targets for the ELTs and ngVLA; and guide the scientific goals of the next-generation missions from X-ray to radio. 
\emph{\textbf{\textcolor{genesisgold}{Now is the time}}} for bold, ambitious surveys to fully complete the picture of how galaxies evolve from first light in the universe to our Milky Way today, accomplishing the first two primary science goals for the JWST mission.

\clearpage

{\fontsize{15}{18}\selectfont\bfseries\color{genesisink}
Authors -- \textcolor{genesisgold}{420 Co-Signers From the Extragalactic Community}\par}
{\large
Alice Shapley\textsuperscript{1}, Steven L. Finkelstein\textsuperscript{2,3}, Rachel S. Somerville\textsuperscript{4}, James Dunlop\textsuperscript{5}, Adriano Fontana\textsuperscript{6}, Pascal A. Oesch\textsuperscript{7,8}, Danielle A. Berg\textsuperscript{2}, Rachel Bezanson\textsuperscript{9}, Maruša Bradač\textsuperscript{10}, Gabriel Brammer\textsuperscript{8}
% ...
}
\vspace{0.5em}

{\small\itshape\textcolor{genesisgray}{%
\textsuperscript{1}University of California, Los Angeles, CA 90095, USA \quad
\textsuperscript{2}Department of Astronomy, The University of Texas at Austin, Austin, TX, USA \quad
\textsuperscript{3}Cosmic Frontier Center, The University of Texas at Austin, Austin, TX, USA \quad
\textsuperscript{4}Center for Computational Astrophysics, Flatiron Institute, New York, NY 10010, USA \quad
\textsuperscript{5}Institute for Astronomy, University of Edinburgh, Royal Observatory, Blackford Hill, Edinburgh EH9 3HJ, UK \quad
\textsuperscript{6}INAF -- Osservatorio Astronomico di Roma, Via Frascati 33, I-00078 Monte Porzio Catone, Italy \quad
\textsuperscript{7}Department of Astronomy, University of Geneva, Chemin Pegasi 51, 1290 Versoix, Switzerland \quad
\textsuperscript{8}Cosmic Dawn Center (DAWN), Niels Bohr Institute, University of Copenhagen, Jagtvej 128, DK-2200 Copenhagen, Denmark \quad
\textsuperscript{9}University of Pittsburgh \quad
\textsuperscript{10}University of Ljubljana, Faculty of Mathematics and Physics, Jadranska ulica 19, SI-1000 Ljubljana, Slovenia \quad
\textsuperscript{11}University of California, Santa Barbara, CA 93106, USA \quad
\textsuperscript{12}Centre for Astrophysics Research, University of Hertfordshire, Hatfield AL10 9AB, UK \quad
\textsuperscript{13}NSF NOIRLab \quad
\textsuperscript{14}Steward Observatory, University of Arizona, 933 North Cherry Avenue, Tucson, AZ 85721, USA \quad
\textsuperscript{15}Department of Astronomy, Graduate School of Science, Kyoto University, Kitashirakawa-Oiwake-cho, Sakyo-ku, Kyoto 606-8502, Japan \quad
\textsuperscript{16}Department of Astronomy \& Astrophysics, University of Chicago, 5640 S Ellis Avenue, Chicago, IL 60637, USA \quad
\textsuperscript{17}Centro de Astrobiología (CAB), CSIC-INTA, Ctra. de Ajalvir km 4, E-28850 Torrejón de Ardoz, Madrid, Spain \quad
\textsuperscript{18}Department of Astronomy and Astrophysics, University of California, Santa Cruz, CA 95064, USA \quad
\textsuperscript{19}Sapienza University of Rome \quad
\textsuperscript{20}Department of Astronomy, Indiana University, 727 East Third Street, Bloomington, IN 47405, USA \quad
\textsuperscript{21}Department of Astronomy, Stockholm University, AlbaNova University Centre, SE-10691 Stockholm, Sweden \quad
\textsuperscript{22}Instituto de Estudios Astrofísicos, Facultad de Ingeniería y Ciencias, Universidad Diego Portales, Av. Ejército 441, Santiago, Chile \quad
\textsuperscript{23}Instituto de Astrofísica e Ciências do Espaço, Portugal \quad
\textsuperscript{24}Faculdade de Ciências da Universidade de Lisboa, Portugal \quad
\textsuperscript{25}SKA Observatory, Jodrell Bank, SK11 9FT, UK \quad
\textsuperscript{26}Institute of Astronomy and Astrophysics, Academia Sinica (ASIAA), Taipei 106319, Taiwan \quad
\textsuperscript{27}University of Nottingham \quad
\textsuperscript{28}Instituto de Astrofísica de Andalucía (IAA-CSIC) \quad
\textsuperscript{29}Institute for Cosmic Ray Research, University of Tokyo \quad
\textsuperscript{30}David A. Dunlap Department of Astronomy and Astrophysics, University of Toronto, Toronto, ON M5S 3H4, Canada \quad
\textsuperscript{31}Universidad de Granada, Departamento de Física Teórica y del Cosmos. Avenida de Fuentenueva S/N, 18071, Granada, Spain \quad
\textsuperscript{32}Centro de Estudios de Física del Cosmos de Aragón (CEFCA), Plaza San Juan 1, 44001 Teruel, Spain \quad
\textsuperscript{33}INAF \quad
\textsuperscript{34}Department of Astronomy, School of Science, The University of Tokyo, 7-3-1, Hongo, Bunkyo, Tokyo, 113-0033, Japan \quad
\textsuperscript{35}Institut d'Astrophysique de Paris, Sorbonne Université, CNRS, UMR 7095, 98 bis bd Arago, 75014 Paris, France \quad
\textsuperscript{36}SURA/NASA Goddard \quad
\textsuperscript{37}DARK, Niels Bohr Institute, University of Copenhagen, Jagtvej 155A, DK-2200 Copenhagen, Denmark \quad
\textsuperscript{38}University of the Pacific \quad
\textsuperscript{39}Centre de Recherche Astrophysique de Lyon (CRAL) \quad
\textsuperscript{40}(1) International Centre for Radio Astronomy Research, University of Western Australia, 7 Fairway, Crawley, WA 6009, Australia (2) Research School of Astronomy and Astrophysics, Australian National University, Weston Creek, ACT 2611, Australia \quad
\textsuperscript{41}Armagh Observatory and Planetarium, College Hill, Armagh, BT61 9DG, N. Ireland, UK \quad
\textsuperscript{42}Università di Bologna, Italy \quad
\textsuperscript{43}University of Massachusetts Amherst, 710 North Pleasant Street, Amherst, MA 01003, USA \quad
\textsuperscript{44}INAF -- Osservatorio di Astrofisica e Scienza dello Spazio di Bologna, via Gobetti 93/3, I-40129 Bologna, Italy \quad
\textsuperscript{45}European Southern Observatory (ESO), Karl-Schwarzschild-Str. 2, 85748 Garching, Germany \quad
\textsuperscript{46}Astronomy Centre, University of Sussex, Falmer, Brighton BN1 9QH, UK \quad
\textsuperscript{47}INAF -- Osservatorio Astronomico di Padova, Vicolo dell'Osservatorio 5, 35122 Padova, Italy \quad
\textsuperscript{48}European Space Agency \quad
\textsuperscript{49}Leiden Observatory, Leiden University, PO Box 9513, NL-2300 RA Leiden, The Netherlands \quad
\textsuperscript{50}Institute for Computational Cosmology, Durham University, South Road, Durham DH1 3LE, UK \quad
\textsuperscript{51}Institute for Theoretical Physics, University of Heidelberg \quad
\textsuperscript{52}School of Earth and Space Exploration, Arizona State University, Tempe, AZ 85287, USA \quad
\textsuperscript{53}Space Telescope Science Institute, 3700 San Martin Drive, Baltimore, MD 21218, USA \quad
\textsuperscript{54}Columbia University \quad
\textsuperscript{55}Department of Physics, University of Oxford, Denys Wilkinson Building, Keble Road, Oxford, OX1 3RH, UK \quad
\textsuperscript{56}Aix Marseille Université, CNRS, CNES, LAM, Marseille, France \quad
\textsuperscript{57}Department of Astronomy and Astrophysics, The Pennsylvania State University, University Park, PA 16802, USA \quad
\textsuperscript{58}Departamento de Fisica y Matematicas, Universidad de Alcala, Campus Universitario, E-28871 Alcala de Henares, Madrid, Spain \quad
\textsuperscript{59}Institute of Astronomy, University of Cambridge, Madingley Road, Cambridge CB3 0HA, UK \quad
\textsuperscript{60}Kavli Institute for Cosmology, University of Cambridge, Madingley Road, Cambridge CB3 0HA, UK \quad
\textsuperscript{61}Dipartimento di Fisica, Sapienza Università di Roma, Piazzale Aldo Moro 2, 00185 Roma, Italy \quad
\textsuperscript{62}Cosmic Dawn Center (DAWN), DTU Space, Technical University of Denmark, 2800 Kgs. Lyngby, Denmark \quad
\textsuperscript{63}Department of Astrophysics, University of Zurich, Winterthurerstrasse 190, CH-8057 Zürich, Switzerland \quad
\textsuperscript{64}Kapteyn Astronomical Institute, University of Groningen \quad
\textsuperscript{65}Scuola Normale Superiore, Piazza dei Cavalieri 7, I-56126 Pisa, Italy \quad
\textsuperscript{66}Dipartimento di Fisica e Astronomia "G. Galilei", Università di Padova, Vicolo dell'Osservatorio 3, 35122 Padova, Italy \quad
\textsuperscript{67}Leiden Observatory, Leiden University, P.O. Box 9513, 2300 RA Leiden, The Netherlands \quad
\textsuperscript{68}Inter-University Centre for Astronomy and Astrophysics, Pune, India \quad
\textsuperscript{69}Department of Physics, The Chinese University of Hong Kong, Shatin, Hong Kong, China \quad
\textsuperscript{70}University of Washington \quad
\textsuperscript{71}University of Michigan \quad
\textsuperscript{72}Sapienza University, Tohoku University \quad
\textsuperscript{73}Jodrell Bank Centre for Astrophysics, University of Manchester, Oxford Road, Manchester M13 9PL, UK \quad
\textsuperscript{74}Departamento de Astronomía, Universidad de Chile, Camino El Observatorio 1515, Las Condes, Santiago, Chile \quad
\textsuperscript{75}Center for Astrophysics \textbar{} Harvard \& Smithsonian, 60 Garden St., Cambridge, MA 02138, USA \quad
\textsuperscript{76}University of Manchester, Oxford Road, Manchester M13 9PL, UK \quad
\textsuperscript{77}Department of Physics \& Astronomy, University of California Irvine, 4129 Reines Hall, Irvine, CA 92697, USA \quad
\textsuperscript{78}Kapteyn Astronomical Institute, University of Groningen, P.O. Box 800, 9700AV Groningen \quad
\textsuperscript{79}Astronomical Institute, Tohoku University, 6-3 Aramaki, Aoba-ku, Sendai, Miyagi 980-8578, Japan \quad
\textsuperscript{80}Instituto de Física y Astronomía, Universidad de Valparaíso, Avda. Gran Bretaña 1111, Valparaíso, Chile \quad
\textsuperscript{81}Millenium Nucleus for Galaxies (MINGAL) \quad
\textsuperscript{82}Instituto de Astrofísica e Ciências do Espaço \quad
\textsuperscript{83}Rochester Institute of Technology \quad
\textsuperscript{84}Universidad Autónoma de Madrid \quad
\textsuperscript{85}Dipartimento di Fisica e Astronomia “Galileo Galilei”, Università \quad
\textsuperscript{86}Centre for Astrophysics and Supercomputing, Swinburne University of Technology \quad
\textsuperscript{87}Department of Physics, 196A Auditorium Road, Unit 3046, University of Connecticut, Storrs, CT 06269, USA and Los Alamos National Laboratory, Los Alamos, NM 87545, USA \quad
\textsuperscript{88}Instituto de Ciencias del Espacio (ICE-CSIC) \quad
\textsuperscript{89}Kapteyn Astronomical Institute, University of Groningen, P.O. Box 800, 9700AV Groningen, The Netherlands \quad
\textsuperscript{90}Institute of Science and Technology Austria (ISTA), Am Campus 1, 3400 Klosterneuburg, Austria \quad
\textsuperscript{91}Instituto de Física de Cantabria (CSIC-UC), Avda. Los Castros s/n, 39005 Santander, Spain \quad
\textsuperscript{92}NSF's National Optical-Infrared Astronomy Research Laboratory, 950 N. Cherry Ave., Tucson, AZ 85719, USA \quad
\textsuperscript{93}University of Hawaii, Hilo, 200 W Kawili St, Hilo, HI 96720, USA \quad
\textsuperscript{94}MIT Kavli Institute for Astrophysics and Space Research, Massachusetts Institute of Technology, Cambridge, MA 02139, USA \quad
\textsuperscript{95}Harvard University \quad
\textsuperscript{96}CEA Paris Saclay \quad
\textsuperscript{97}University College London \quad
\textsuperscript{98}IPAC, California Institute of Technology, 1200 E. California Blvd. Pasadena, CA 91125, USA \quad
\textsuperscript{99}Institute of Physics, Laboratory for Galaxy Evolution and Spectral Modelling, EPFL, Observatoire de Sauverny, 1290 Versoix, Switzerland \quad
\textsuperscript{100}International Gemini Observatory/NSF NOIRLab, 670 N. A'ohoku Place, Hilo, HI 96720, USA \quad
\textsuperscript{101}Instituto de Astrofísica de Canarias, c/ Vía Láctea s/n, E-38205 La Laguna, Tenerife, Spain \quad
\textsuperscript{102}Uppsala University \quad
\textsuperscript{103}Department of Astronomy, Astrophysics and Space Engineering, Indian Institute of Technology Indore, Simrol, MP 453552, India \quad
\textsuperscript{104}Kavli IPMU (WPI), UTIAS, The University of Tokyo, Kashiwa, Chiba 277-8583, Japan \quad
\textsuperscript{105}Rutgers University \quad
\textsuperscript{106}Cresci \quad
\textsuperscript{107}JWST Australian Data Centre, Swinburne University of Technology \quad
\textsuperscript{108}Universitaet Heidelberg, Zentrum fuer Astronomie, Institut fuer Theoretische Astrophysik, Albert-Ueberle-Str. 2, D-69120 Heidelberg, Germany \quad
\textsuperscript{109}Fermilab \quad
\textsuperscript{110}Instituto de Astrofísica de Andalucía-CSIC (Granada, Spain) \quad
\textsuperscript{111}Department of Astrophysical Sciences, Princeton University, Princeton, NJ 08544, USA \quad
\textsuperscript{112}University of Missouri, Columbia \quad
\textsuperscript{113}Institute for Astronomy, University of Hawaii \quad
\textsuperscript{114}Department of Physics and Astronomy, University of California Riverside, Riverside, CA 92521, USA \quad
\textsuperscript{115}Center for Space Sciences and Technology, UMBC, 5523 Research Park Dr, Baltimore, MD 21228 USA \quad
\textsuperscript{116}Astrophysics Science Division, NASA Goddard Space Flight Center, 8800 Greenbelt Rd, Greenbelt, MD 20771, USA \quad
\textsuperscript{117}University of Cambridge \quad
\textsuperscript{118}U.S. National Science Foundation (NSF), Alexandria, VA 22314, USA \quad
\textsuperscript{119}Department of Physics, University of California, Merced, 5200 Lake Road, Merced, CA 92543, USA \quad
\textsuperscript{120}Institute of Physics, Lab for galaxy evolution and spectral modelling, Observatory of Sauverny, Chemin Pegasi 51, Versoix, Switzerland \quad
\textsuperscript{121}Department of Physics, University of Louisville, Natural Science Building 102, 40292 KY Louisville, USA \quad
\textsuperscript{122}Departamento de Astrofísica, Universidad de La Laguna, E-38205 La Laguna, Tenerife, Spain \quad
\textsuperscript{123}Department of Astrophysics, University of Vienna, Tuerkenschanzstr. 17, 1180 Vienna, Austria \quad
\textsuperscript{124}Laboratory for Multiwavelength Astrophysics, School of Physics and Astronomy, Rochester Institute of Technology, 84 Lomb Memorial Drive, Rochester, NY 14623, USA \quad
\textsuperscript{125}Kavli Institute for Astronomy and Astrophysics, Peking University, Beijing 100871, People’s Republic of China \quad
\textsuperscript{126}Waseda University, Tokyo, Japan \quad
\textsuperscript{127}Institute for Astronomy, UH Mānoa \quad
\textsuperscript{128}National Astronomical Observatory of Japan, 2-21-1 Osawa, Mitaka, Tokyo 181-8588, Japan \quad
\textsuperscript{129}Univ Toulouse, CNES, CNRS, IRAP, Toulouse, France \quad
\textsuperscript{130}Institut d’Astrophysique Spatiale, Université Paris-Saclay, 91405 Orsay, France \quad
\textsuperscript{131}Cavendish Laboratory, University of Cambridge, 19 JJ Thomson Avenue, Cambridge CB3 0HE, UK \quad
\textsuperscript{132}Department of Physics and Astronomy, University of California, Davis, 1 Shields Ave, Davis, CA 95616, USA \quad
\textsuperscript{133}The University of Maine \quad
\textsuperscript{134}Chungbuk National University, Cheongju, 28644, Republic of Korea \quad
\textsuperscript{135}Max Planck Institute for Astrophysics, Karl-Schwarzschild-Str. 1, 85748 Garching, Germany \quad
\textsuperscript{136}Department of Astronomy \& the DiRAC Institute, University of Washington, Physics-Astronomy Building, Box 351580, Seattle, WA 98195-1700, USA \quad
\textsuperscript{137}Yonsei University \quad
\textsuperscript{138}Department of Physics and Astronomy, Colby College, Waterville, ME 04901, USA \quad
\textsuperscript{139}Tokyo University of Technology \quad
\textsuperscript{140}Rensselaer Polytechnic Institute \quad
\textsuperscript{141}Leiden Observatory \quad
\textsuperscript{142}The University of Tokyo, Kashiwa, Chiba 277-8582, Japan \quad
\textsuperscript{143}Telespazio UK for ESA, ESAC, ESAC, Camino Bajo del Castillo s/n, E-28692 Villanueva de la Cañada, Madrid, Spain \quad
\textsuperscript{144}NASA-Goddard \quad
\textsuperscript{145}SISSA, Via Bonomea 265, 34136 Trieste, Italy \quad
\textsuperscript{146}Physics Department, College of Science, United Arab Emirates University, PO Box 15551, Al-Ain, UAE \quad
\textsuperscript{147}Kavli IPMU (University of Tokyo) \quad
\textsuperscript{148}Massachusetts Institute of Technology \quad
\textsuperscript{149}School of Physics and Technology, Wuhan University, Wuhan 430072, People’s Republic of China. \quad
\textsuperscript{150}Department of Astronomy, Tsinghua University, Beijing 100084, People's Republic of China \quad
\textsuperscript{151}Como Lake Center for Astrophysics, University of Insubria \quad
\textsuperscript{152}INAF -- Italian National Institute of Astrophysics, Astronomical Observatory of Trieste \quad
\textsuperscript{153}INAF -- Osservatorio Astrofisico di Arcetri, Largo E. Fermi 5, 50125 Firenze, Italy \quad
\textsuperscript{154}Homer L. Dodge Department of Physics and Astronomy, University of Oklahoma \quad
\textsuperscript{155}Department of Astrophysics, University of Zurich, Winterthurerstrasse 190, 8057, Zurich, Switzerland \quad
\textsuperscript{156}University of Catania \quad
\textsuperscript{157}Astronomical Institute, Graduate School of Science, Tohoku University, 6–3 Aoba, Sendai 980-8578, Japan \quad
\textsuperscript{158}Dipartimento di Fisica e Astronomia, Alma Mater Studiorum Università di Bologna \quad
\textsuperscript{159}The University of Osaka \quad
\textsuperscript{160}Sydney Institute for Astronomy, School of Physics, University of Sydney, NSW 2006, Australia \quad
\textsuperscript{161}University of Florida \quad
\textsuperscript{162}Universität Heidelberg, Institut für Theoretische Astrophysik, Zentrum für Astronomie, AU-Str. 2, 69120 Heidelberg, Germany \quad
\textsuperscript{163}Observatoire astronomique de Strasbourg, Université de Strasbourg, CNRS UMR 7550, 11 rue de l'Université, 67000 Strasbourg, France \quad
\textsuperscript{164}Tohoku University \quad
\textsuperscript{165}Institute for Cosmic Ray Research, The University of Tokyo, 5-1-5 Kashiwanoha, Kashiwa, Chiba 277-8582, Japan \quad
\textsuperscript{166}Astrophysics Research Institute, Liverpool John Moores University, 146 Brownlow Hill, Liverpool, L3 5RF, UK \quad
\textsuperscript{167}The Observatories of the Carnegie Institution for Science \quad
\textsuperscript{168}Department of Astronomy, University of Illinois Urbana-Champaign \quad
\textsuperscript{169}Department of Physics and Astronomy, Tufts University, 574 Boston Ave., Medford, MA 02155, USA \quad
\textsuperscript{170}Texas A\&M University \quad
\textsuperscript{171}Kavli Institute for Astronomy and Astrophysics (KIAA), Peking University, Beijing 100871, China \quad
\textsuperscript{172}Dunlap Institute for Astronomy \& Astrophysics, University of Toronto, Toronto, ON, Canada \quad
\textsuperscript{173}Dipartimento di Fisica e Astronomia, Università di Bologna, Via Gobetti 93/2, 40129 Bologna, Italy \quad
\textsuperscript{174}Maynooth University \quad
\textsuperscript{175}Konkoly Observatory \quad
\textsuperscript{176}Instituto Galego de Física de Altas Enerxías, Universidade de Santiago de Compostela, 15782 Santiago de Compostela, Galicia, Spain \quad
\textsuperscript{177}University of Bath \quad
\textsuperscript{178}Department of Astronomy, University of Maryland, College Park, MD 20742, USA \quad
\textsuperscript{179}University of Southampton \quad
\textsuperscript{180}Dipartimento di Fisica e Astronomia, Universita' di Padova \quad
\textsuperscript{181}Kavli Institute for Cosmological Physics, University of Chicago, Chicago IL 60637, USA \quad
\textsuperscript{182}University of Oxford \quad
\textsuperscript{183}National Centre for Radio Astrophysics, TIFR, Pune, India \& International Centre for Theoretical Sciences, TIFR, Bengaluru, India \quad
\textsuperscript{184}IUCAA \quad
\textsuperscript{185}School of Mathematical and Physical Sciences, Macquarie University, Sydney, NSW 2109, Australia \quad
\textsuperscript{186}Florence University, Department of Physics and Astronomy, Largo Enrico Fermi, 50124, Florence, Italy \quad
\textsuperscript{187}Department of Astronomy, National Research Institute of Astronomy and Geophysics (NRIAG), Cairo, 11421, Egypt \quad
\textsuperscript{188}Department of Physics and Astronomy, University of Kentucky, 505 Rose Street, Lexington, KY 40506, USA \quad
\textsuperscript{189}Hamburg Observatory, University of Hamburg, Gojenbergsweg 112, D-21029 Hamburg, Germany \quad
\textsuperscript{190}School of Physics and Astronomy, University of Southampton, Highfield, Southampton, SO17 1BJ, UK \quad
\textsuperscript{191}University of Illinois at Urbana-Champaign \quad
\textsuperscript{192}MIT \quad
\textsuperscript{193}Departamento de Astronomía, Universidad de Chile, Camino el Observatorio 1515, Las Condes, Santiago, Chile \quad
\textsuperscript{194}Providence College \quad
\textsuperscript{195}Liverpool John Moores University \quad
\textsuperscript{196}University of Warwick, Department of Physics, Gibbet Hill Road, Coventry, CV4 7AL, UK \quad
\textsuperscript{197}Universitat de València \quad
\textsuperscript{198}NASA's Goddard Space Flight Center \quad
\textsuperscript{199}Department of Natural Sciences, Faculty of Science and Engineering, Tokyo City University, 1-28-1 Tamazutsumi, Setagaya, Tokyo 158–8557, Japan \quad
\textsuperscript{200}Northwestern University \quad
\textsuperscript{201}Nagoya University, Furocho, Chikusa, Nagoya 464-8602, Japan \quad
\textsuperscript{202}Kavli IPMU \quad
\textsuperscript{203}School of Astronomy and Space Science, Nanjing University, Nanjing, Jiangsu 210023, China \quad
\textsuperscript{204}Aix Marseille Univ, CNRS, CNES, LAM, Marseille, France \quad
\textsuperscript{205}Ritsumeikan University, Kusatsu, Shiga, Japan \quad
\textsuperscript{206}MIT Kavli Institute \quad
\textsuperscript{207}Tel Aviv University \quad
\textsuperscript{208}School of Physics, The University of Melbourne, VIC 3010, Australia \quad
\textsuperscript{209}Department of Physics, University of Connecticut, 2152 Hillside Road, Unit 3046, Storrs, CT 06269, USA \quad
\textsuperscript{210}Kavli Institute for the Physics and Mathematics of the Universe (WPI), The University of Tokyo, Kashiwa, Chiba 277-8583, Japan \quad
\textsuperscript{211}Max Planck Institute for Extraterrestrial Physics, Giessenbachstrasse 1, 85748 Garching, Germany \quad
\textsuperscript{212}Department of Physics and Kavli Institute for Astrophysics and Space Research, Massachusetts Institute of Technology, Cambridge, MA 02139, USA \quad
\textsuperscript{213}Max Planck Institute for Astronomy, Königstuhl 17, 69117 Heidelberg, Germany \quad
\textsuperscript{214}Department of Astronomy, University of Michigan, 1085 S. University Ave., Ann Arbor, MI 48109, USA \quad
\textsuperscript{215}Brinson Prize Fellow, MIT Kavli Institute for Astrophysics and Space Research \quad
\textsuperscript{216}Department of Physics \& Astronomy, Ghent University, Belgium \quad
\textsuperscript{217}Georgia Institute of Technology \quad
\textsuperscript{218}Universidad de Chile \quad
\textsuperscript{219}Kavli Institute for the Physics and Mathematics of the Universe (WPI), The University of Tokyo Institutes for Advanced Study, The University of Tokyo, Kashiwa, Chiba 277-8583, Japan \quad
\textsuperscript{220}Graduate Institute of Astrophysics, National Taiwan University, Taipei, Taiwan \quad
\textsuperscript{221}University of Missouri-Columbia \quad
\textsuperscript{222}Department of Physics, Graduate School of Science, The University of Tokyo, 7-3-1 Hongo, Bunkyo, Tokyo 113-0033, Japan \quad
\textsuperscript{223}Nanjing Institute of Astronomical Optics \& Technology, Chinese Academy of Sciences \quad
\textsuperscript{224}National Tsing Hua University - Taiwan \quad
\textsuperscript{225}Observational Astrophysics, Department of Physics and Astronomy, Uppsala University, Box 524, SE-751 20 Uppsala, Sweden \quad
\textsuperscript{226}Dipartimento di Fisica, Sapienza, Università di Roma, Piazzale Aldo Moro 5, IT-00185 Roma, Italy \quad
\textsuperscript{227}Department of Astronomy, School of Physics, Huazhong University of Science and Technology, Wuhan, Hubei 430074, China \quad
\textsuperscript{228}Laboratoire d'Astrophysique de Marseille \quad
\textsuperscript{229}Canadian Institute for Theoretical Astrophysics \quad
\textsuperscript{230}Kavli Institute for Particle Astrophysics \& Cosmology, Stanford University, 452 Lomita Mall, Stanford, CA 94305, USA \quad
\textsuperscript{231}The University of British Columbia \quad
\textsuperscript{232}University of Illinois Urbana-Champaign \quad
\textsuperscript{233}Department of Physics, Ben-Gurion University of the Negev, P.O. Box 653, Be'er-Sheva 84105, Israel}}
\vspace{0.4em}

\bibliography{genesis,rss}{}
\bibliographystyle{aasjournalv7.1}

\end{document}